\RequirePackage{lineno}
\documentclass[prd,twocolumn,amsmath,amssymb]{revtex4}
\usepackage{graphicx}
\usepackage{dcolumn}
\usepackage{bm}
\usepackage{epsfig}
\usepackage{overpic}
\usepackage{verbatim}
\newsavebox{\tablebox}
\usepackage{epstopdf}
\usepackage[colorlinks,linkcolor=blue,anchorcolor=red,citecolor=blue,breaklinks=true]{hyperref}

\usepackage{rotating}

\newcommand{\AllDneChannels}{D^{\pm}\to n(\bar{n})e^{\pm}}
\newcommand{\Dptonbare}{D^{+}\to \bar{n}e^{+}}
\newcommand{\Dptone}{D^{+}\to ne^{+}}
\newcommand{\Dmtonbare}{D^{-}\to \bar{n}e^{-}}
\newcommand{\Dmtone}{D^{-}\to ne^{-}}
\newcommand{\CombChannelsA}{D^{+(-)}\to \bar{n}(n)e^{+(-)}}
\newcommand{\CombChannelsB}{D^{+(-)}\to n(\bar{n})e^{+(-)}}
\newcommand{\Dp}{D^{+}}
\newcommand{\Dm}{D^{-}}
\newcommand{\Dpm}{D^{\pm}}
\newcommand{\Ks}{K_{S}^{0}}
\newcommand{\Kl}{K_{L}^{0}}
\newcommand{\piz}{\pi^{0}}
\newcommand{\DptoKpipi}{D^{+}\to K^-\pi^{+}\pi^{+}}
\newcommand{\DptoKpipipiz}{D^{+}\to K^-\pi^{+}\pi^{+}\pi^{0}}
\newcommand{\DptoKspi}{D^{+}\to K_S^{0}\pi^{+}}
\newcommand{\DptoKspipiz}{D^{+}\to K_S^{0}\pi^{+}\pi^{0}}
\newcommand{\DptoKspipipi}{D^{+}\to K_S^{0}\pi^{+}\pi^{+}\pi^{-}}
\newcommand{\DptoKKpi}{D^{+}\to K^{+}K^{-}\pi^{+}}
\newcommand{\gevcc}{{\rm GeV}/c^{2}}

\newcommand{\BR}{{\cal B}}
\newcommand{\chisq}{\chi^{2}}

\newcommand{\dataIntLumi}{\rm 2.93~fb^{-1}}
\newcommand{\BrDataCombA}{1.43\times10^{-5}}
\newcommand{\BrDataCombB}{2.91\times10^{-5}}

\newcommand {\NIMA}    {Nucl.{} Instrum.{} Meth.{} A }
\newcommand {\PRD}     {Phys.{} Rev.{} D }

\newcommand {\PRL}     {Phys.{} Rev.{} Lett.{} }
\newcommand {\PLB}     {Phys.{} Lett.{} B }

\newcommand {\CPC} {Chin.{} Phys.{} C }

\begin{document}

\normalsize
\parskip=5pt plus 1pt minus 1pt

\title{\boldmath Search for baryon and lepton number violation decay $\AllDneChannels$}
\author{
\begin{small}
\begin{center}
M.~Ablikim$^{1}$, M.~N.~Achasov$^{11,b}$, P.~Adlarson$^{70}$, M.~Albrecht$^{4}$, R.~Aliberti$^{31}$, A.~Amoroso$^{69A,69C}$, M.~R.~An$^{35}$, Q.~An$^{66,53}$, X.~H.~Bai$^{61}$, Y.~Bai$^{52}$
% ...
\\
\vspace{0.2cm}
(BESIII Collaboration)\\
\vspace{0.2cm} {\it
$^{1}$ Institute of High Energy Physics, Beijing 100049, People's Republic of China\\
% ...
$^{4}$ Bochum Ruhr-University, D-44780 Bochum, Germany\\
% ...
$^{11}$ G.I. Budker Institute of Nuclear Physics SB RAS (BINP), Novosibirsk 630090, Russia\\
% ...
$^{31}$ Johannes Gutenberg University of Mainz, Johann-Joachim-Becher-Weg 45, D-55099 Mainz, Germany\\
% ...
$^{35}$ Liaoning Normal University, Dalian 116029, People's Republic of China\\
% ...
$^{52}$ Southeast University, Nanjing 211100, People's Republic of China\\
$^{53}$ State Key Laboratory of Particle Detection and Electronics, Beijing 100049, Hefei 230026, People's Republic of China\\
% ...
$^{61}$ University of Jinan, Jinan 250022, People's Republic of China\\
% ...
$^{66}$ University of Science and Technology of China, Hefei 230026, People's Republic of China\\
% ...
$^{69}$ University of Turin and INFN, (A)University of Turin, I-10125, Turin, Italy; (B)University of Eastern Piedmont, I-15121, Alessandria, Italy; (C)INFN, I-10125, Turin, Italy\\
$^{70}$ Uppsala University, Box 516, SE-75120 Uppsala, Sweden\\
% ...
$^{b}$ Also at the Novosibirsk State University, Novosibirsk, 630090, Russia\\
% ...
}
\end{center}
\end{small}
}

\begin{abstract}
Using a data set of electron-positron collisions corresponding to an integrated luminosity of $\dataIntLumi$ taken with the BESIII detector at a center-of-mass energy of 3.773~GeV, a search for the baryon ($B$) and lepton ($L$) number violating decays $\AllDneChannels$ is performed. No signal is observed and the upper limits on the branching fractions at the $90\%$ confidence level are set to be $\BrDataCombA$ for the decays $\CombChannelsA$ with $\Delta|B-L|=0$, and $\BrDataCombB$ for the decays $\CombChannelsB$ with  $\Delta|B-L|=2$ , where $\Delta|B-L|$ denotes the change in the difference between baryon and lepton numbers.

\end{abstract}

\maketitle

\section{INTRODUCTION}

\begin{figure}[!htp]
  \centering
  \mbox{
	\begin{overpic}[width=0.15\textwidth,angle=270]{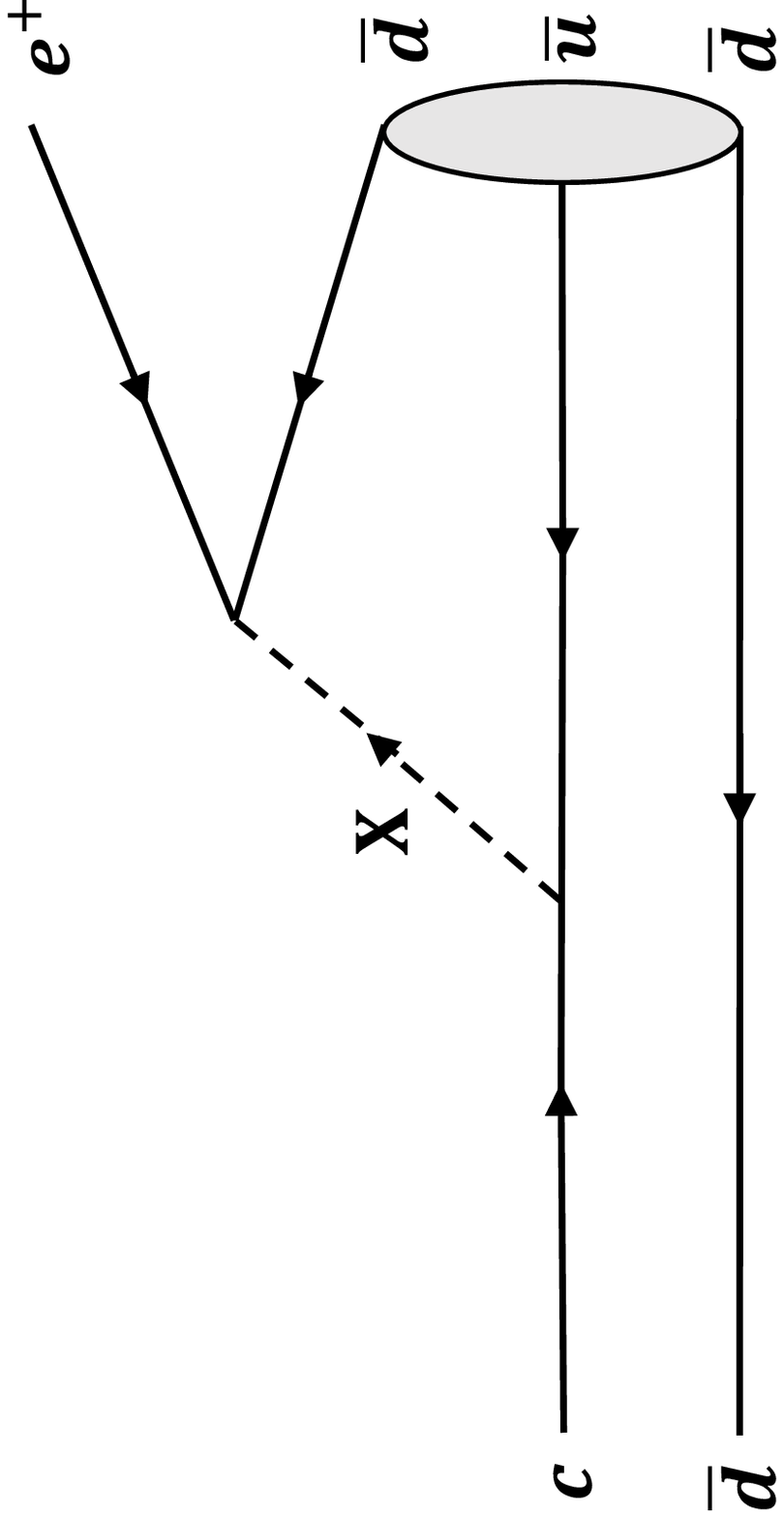}\put(50,0){$(a)$}\end{overpic}
	\begin{overpic}[width=0.15\textwidth,angle=270]{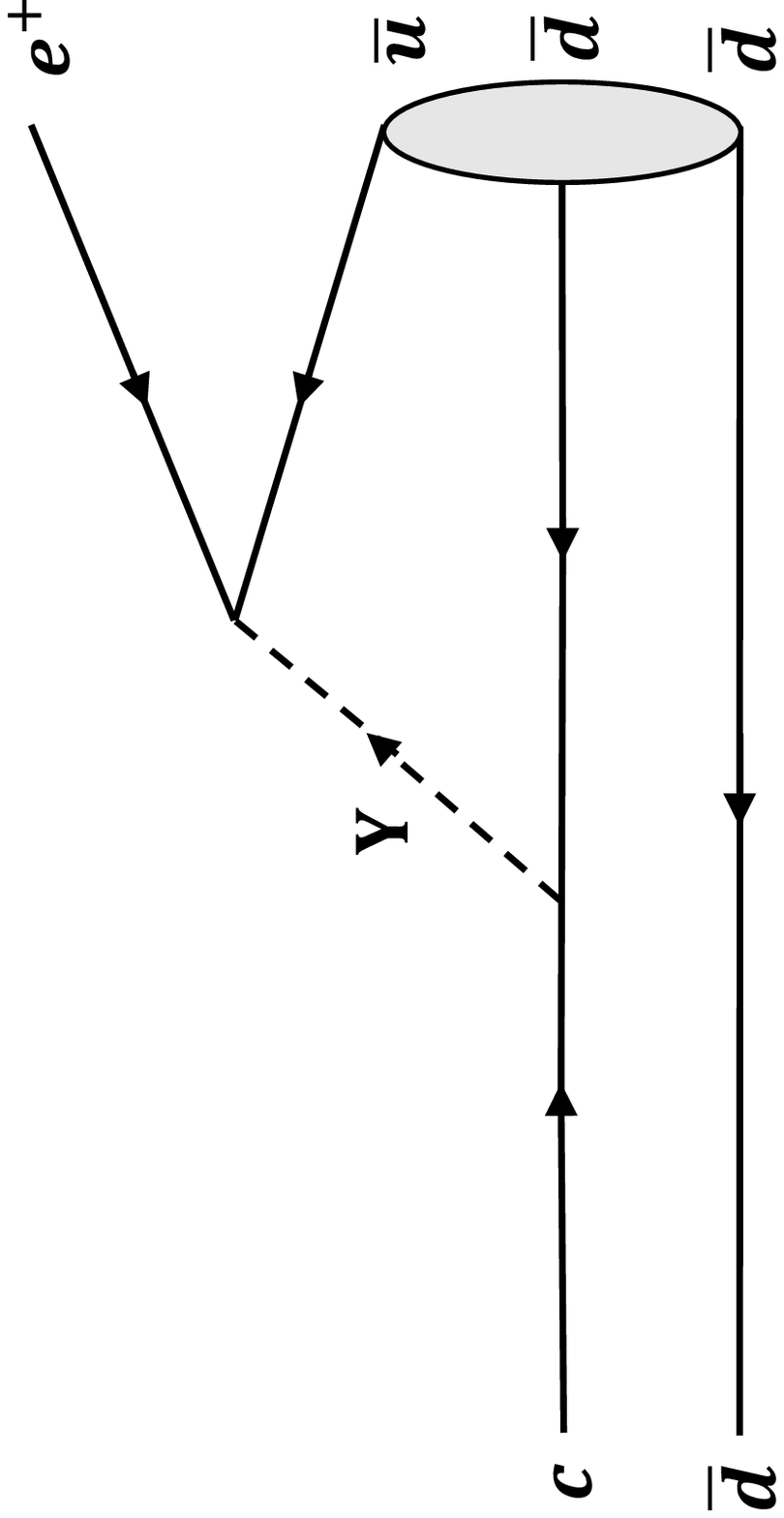}\put(50,0){$(b)$}\end{overpic}
  }
  \begin{overpic}[width=0.2\textwidth,angle=0]{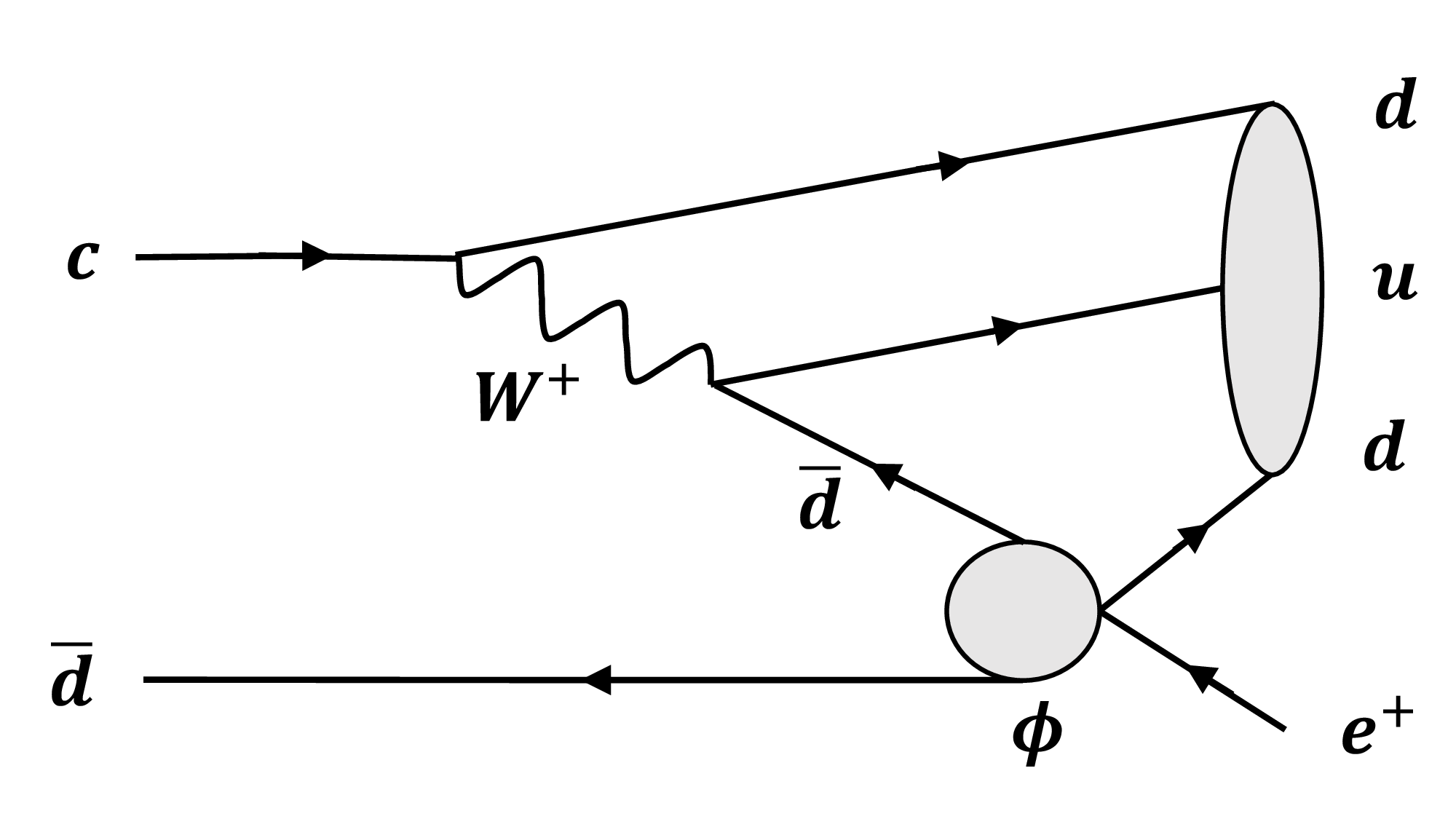}\put(50,0){$(c)$}\end{overpic}
  \caption{
     Feynman diagrams for $\Dptonbare$ with heavy gauge bosons $X$ (a) and $Y$ (b), and  $\Dptone$ with elementary scalar fields $\phi$ (c).
  }
  \label{fig:FeynmanDiagram}
\end{figure}

The search for new physics beyond the Standard Model (SM) is one of the major goals of particle physics.
The matter-antimatter asymmetry of the universe is one prominent observations that cannot be explained within the SM, and as such is a serious challenge to our understanding of nature. 
This asymmetry suggests the existence of baryon number violation (BNV)~\cite{sakharov}.
While proton decay has been searched for decades but not yet observed, 
the search for decays of heavy mesons and baryons that are forbidden in the SM can provide an alternative probe to search for BNV.
In most  grand unified theories (GUTs)~\cite{GUTs1,GUTs2,GUTs3,GUTs4,GUTs5} and some SM extension models~\cite{SUSY1,SUSY2}, baryon-number and lepton-number violation (LNV) is allowed, but the difference of baryon and lepton numbers is conserved  ($\Delta|B-L|=0$).
Dimension-six operators allow processes with $\Delta|B-L|=0$ to proceed, mediated by heavy gauge bosons $X$ with charge $\frac{4}{3}$ or $Y$ with charge $\frac{1}{3}$, as shown in Feynman diagrams in Fig.~\ref{fig:FeynmanDiagram} (a) and (b) for $D$ meson decays.
Furthermore, there is another BNV process possible under dimension-seven operators, mediated by an elementary scalar field $\phi$, as shown in Fig.~\ref{fig:FeynmanDiagram} (c). In this process, the difference of baryon and lepton number is changed by 2 units ($\Delta|B-L|=2$).
Reference~\cite{B-LInProtonDecay} argues that the decay amplitudes of these two kinds of BNV processes are expected to be of comparable strength.
Thus, experimental searches for these BNV decays probe new physics effects and test different models beyond the SM.

The CLEO, BABAR, and CLAS experiments searched for BNV processes in $D$, $B$ meson and hyperon decays ~\cite{DtopebyCLEO,BNVInBbyBARBAR,BNVInHyperonbyCLAS}, respectively, without finding evidence of a signal. Upper limits (ULs) were set on the decay branching fractions in the range of $10^{-5}\sim 10^{-8}$  at the 90\% confidence level (CL). 
Recently, 
the BESIII experiment searched for $D^+$ meson decays to a hyperon and an electron, \emph{i.e.} $D^+\to \bar{\Lambda} (\bar{\Sigma^0}) e^+$ with $\Delta (B-L)=0$ and  $D^+\to {\Lambda} ({\Sigma}^0) e^+$ with $\Delta (B-L)=2$. No  signal was found and ULs of around  $10^{-6}$ were set on the decay branching fractions at the 90\% CL~\cite{DtoLambdae}. 
It is natural to extend the search to $D^+$ meson decays to a (anti-)neutron and electron pair.
A higher generation SUSY model~\cite{SUSYmodel} predicts the branching fraction of $D^{0}\to\bar{p} \ell^{+}$ ($\ell^+ = e^+,\mu^+$) to be less than $4.0\times 10^{-39}$, thus the decay $D^{+}\to \bar{n}\ell^{+}$ is also expected to be of a comparable magnitude because it differs only by the change of a spectator quark.
 
 In this paper, we report the first search for the  BNV process $\CombChannelsA$ with $\Delta|B-L|=0$,  and $\CombChannelsB$  with $\Delta|B-L|=2$  by using  $\dataIntLumi$ of electron-positron collision data taken at a center-of-mass energy of $\sqrt{s}=3.773$~GeV. 
Throughout this paper, the presence of charge-conjugated processes are implied unless explicitly stated otherwise.

\section{BESIII DETECTOR AND MONTE CARLO SIMULATION}
The BESIII detector is a magnetic spectrometer~\cite{BESIIIDetector} located at the Beijing Electron Positron Collider (BEPCII)~\cite{bepcii}. The cylindrical core of the BESIII detector consists of a helium-based multilayer drift chamber (MDC), a plastic scintillator time-of-flight system (TOF), and a CsI(Tl) electromagnetic calorimeter (EMC), which are all enclosed in a superconducting solenoidal magnet providing a 1.0~T magnetic field. The solenoid is supported by an octagonal flux-return yoke with resistive plate counter muon identifier modules interleaved with steel. 
The acceptance of charged particles and photons is 93\% over $4\pi$ solid angle. The charged-particle momentum resolution at $1~{\rm GeV}/c$ is $0.5\%$, and the specific energy loss (d$E$/d$x$) resolution is $6\%$ for the electrons from Bhabha scattering. The EMC measures photon energies with a resolution of $2.5\%$ ($5\%$) at $1$~GeV in the barrel (end-cap) region. The TOF measures flight time of charged particle with a resolution 68~ps in the barrel region, and 110~ps in the end-cap region when the data sets in this analysis were collected. 

Monte Carlo (MC) simulation  samples, generated with a {\sc geant4}-based~\cite{geant4} package~\cite{BESIIIMC}, including the geometric and material description of the BESIII detector, are used to determine the detection efficiency, optimize the selection criteria and estimate the backgrounds. The analysis is performed in the framework of the BESIII Offline Software System~\cite{BOSS}, which takes care of the detector calibration, event reconstruction and date storage. The simulation includes the beam energy spread and initial state radiation (ISR) in the $e^{+}e^{-}$ annihilations modeled with the generator {\sc kkmc}~\cite{KKMC}. The inclusive MC samples contain the production of $D\bar{D}$ pairs, the non-$D\bar{D}$ decays of the $\psi(3770)$, the ISR production of the $J/\psi$ and $\psi(3686)$ states, and  continuum processes, in which the known decay modes are modeled with {\sc evtgen}~\cite{EVTGEN} using branching fractions taken from the Particle Data Group (PDG)~\cite{PDG}, and the remaining unknown decays from charmonium states are modeled with {\sc lundcharm}~\cite{LUNDCHARM1,LUNDCHARM2}. The final-state radiation from charged particles is incorporated using {\sc photos}~\cite{PHOTOS}. 
In the signal MC sample,  $\Dp\Dm$ pairs are generated by the VSS model from {\sc evtgen}~\cite{EVTGEN} and the signal process is generated with a uniform momentum distribution in the phase space (PHSP) according to the conservation of angular momentum. 

\section{Data Analysis}
\subsection{Analysis Method}
At $\sqrt{s}=3.773$ GeV, $D^{\pm}$ mesons are produced in pairs without the presence of any additional fragmentation particles. This property provides an ideal environment for investigating $D^\pm$ meson decays with a double-tag (DT) method~\cite{DoubleTag}. In this approach, the single-tag (ST) $\Dp$ meson is reconstructed in six hadronic decay modes $\DptoKpipi$, $\DptoKpipipiz$, $\DptoKspi$, $\DptoKspipiz$, $\DptoKspipipi$ and $\DptoKKpi$, all of which have relatively large branching fractions and low background contamination. 
The DTs are then formed by reconstructing the other charm meson in the event in its decay to the signal mode.
The decay branching fractions of the four signal modes \mbox{($\CombChannelsA$ and $\CombChannelsB$)} are determined independently by
\begin{equation}
  \label{eq:BR}
  \BR_{\rm sig} = N_{\rm DT}/({N_{\rm ST}^{\rm tot} \cdot \epsilon_{\rm sig}}),
\end{equation}
where $N_{\rm ST}^{\rm tot}$ and $N_{\rm DT}$ are respectively the yields of ST and DT events in data summed over all ST decay modes. 
The effective signal detection efficiency in the presence of the ST $\Dpm$ meson is calculated by \mbox{$\epsilon_{\rm sig}=\Sigma_{i}[(\epsilon_{\rm DT}^{i} \cdot N_{\rm ST}^{i})/(\epsilon_{\rm ST}^{i} \cdot N_{\rm ST}^{\rm tot})]$}, where $\epsilon^{i}_{\rm ST}$ and $\epsilon^{i}_{\rm DT}$ are the corresponding detection efficiencies of the ST and DT method for the $i^{th}$ tag mode, respectively, and $i$ sums over all ST decay modes. 

To avoid possible bias, a blind analysis technique is performed in which the data in the interesting phase space region are viewed only after the analysis strategy is validated with MC simulation or data in the control region and then fixed.

\subsection{Event Selection}
Charged particles, including kaon, pion and electron/positron candidates, are reconstructed from the hit information in the MDC. 
The charged tracks, apart from in the case of pions from  the decays of candidate $\Ks$ mesons, are required to have a distance of closest approach to the interaction point (IP) within $\pm$10~cm in the beam direction and 1~cm in the plane perpendicular to the beam.
The polar angle $\theta$ of charged tracks with respect to the $z$-axis of the MDC must satisfy $|\!\cos\!\theta| < 0.93$. 
Particle identification (PID), based on the information from the d$E$/d$x$ and TOF measurements, is applied to the charged tracks and the CLs for the kaon and pion hypotheses ($CL_{K,\pi}$) are calculated.
A kaon is identified by requiring $CL_{K}>CL_{\pi}$, and a pion by requiring $CL_{\pi}>CL_{K}$.
To identify electrons, the deposited energy in the EMC is utilized, in addition to the d$E$/d$x$ and TOF information,  and the CLs are calculated for the  electron, pion, kaon, and proton hypotheses ($CL_{e,\pi,K, p}$), individually. 
An electron is identified by requiring  $CL_{e}/(CL_{e}+CL_{\pi}+CL_{K}+CL_{p})>0.8$. 

Photon candidates are reconstructed from EMC showers and are required to have energy greater than 25~MeV in the barrel region ($|\!\cos\!\theta|<0.8$), and 50~MeV in the end-cap region ($0.86<|\!\cos\!\theta|<0.92$).
To suppress showers from electronic noise and those unrelated to the event under analysis, the EMC shower time is required to be within 700~ns of the start time of the event.
The minimum opening angle between the photon candidate and all charged tracks is required to be greater than $10^{\circ}$ to avoid contamination from charged tracks showering in the EMC detector.
The $\piz$ candidates are reconstructed from photon pairs by requiring the invariant mass ($M_{\gamma\gamma}$) to be in the range (0.115,0.150)~$\gevcc$. To improve the kinematic resolution, a kinematic fit constraining $M_{\gamma\gamma}$ to the nominal $\piz$ mass~\cite{PDG} is applied to the $\piz$ candidate.
The kinematic variables after the kinematic fit are used in the subsequent analysis.

The $\Ks$ candidates are reconstructed from two charged tracks with opposite charges, polar angle in the range $|\!\cos\!\theta| < 0.93$ and points of closest approach to the IP within $\pm$20~cm along  the beam direction. No requirement on the distance of closest approach in the plane perpendicular to the $z$ direction is applied. A vertex fit is performed on the two tracks on the assumption that they are pions from a common decay point. A further secondary vertex fit, which constrains the $\Ks$ to come from the beamspot, is applied to suppress background with the requirement $L/\sigma_{L}>2$, where $L$ is the decay length, defined as the distance between the primary and secondary vertexes, and $\sigma_{L}$ is the corresponding resolution. The invariant mass ($M_{\pi\pi}$) must satisfy $|M_{\pi\pi}-M_{\Ks}|<0.012\gevcc$, where $M_{\Ks}$ is the known mass of the 
$\Ks$ meson~\cite{PDG}.

\subsection{Single-Tag Events}
The ST $D^\pm$ mesons are reconstructed in the six ST hadronic-decay modes and separated from background using two variables: the energy difference $\Delta E=E_{D}-E_{\rm beam}$ and the beam constrained mass $M_{\rm BC}=\sqrt{E_{\rm beam}-p_{D}^2}$, where $E_{\rm beam}$ is beam energy, $E_{D}$ and $p_{D}$ are the energy and momentum of the ST $D^\pm$ meson candidates in the rest frame of the $e^{+}e^{-}$ system, respectively. When multiple candidates for a specific ST mode are found, the one with minimum $|\Delta E|$ is retained. 
The ST candidate events are further required to have $\Delta E$ within $(-55,+40)$~MeV for ST modes including a $\piz$, and within $(-25,+25)$~MeV otherwise. 

\begin{figure}[!htp]
  \centering
  \mbox{
	\begin{overpic}[width=0.45\textwidth]{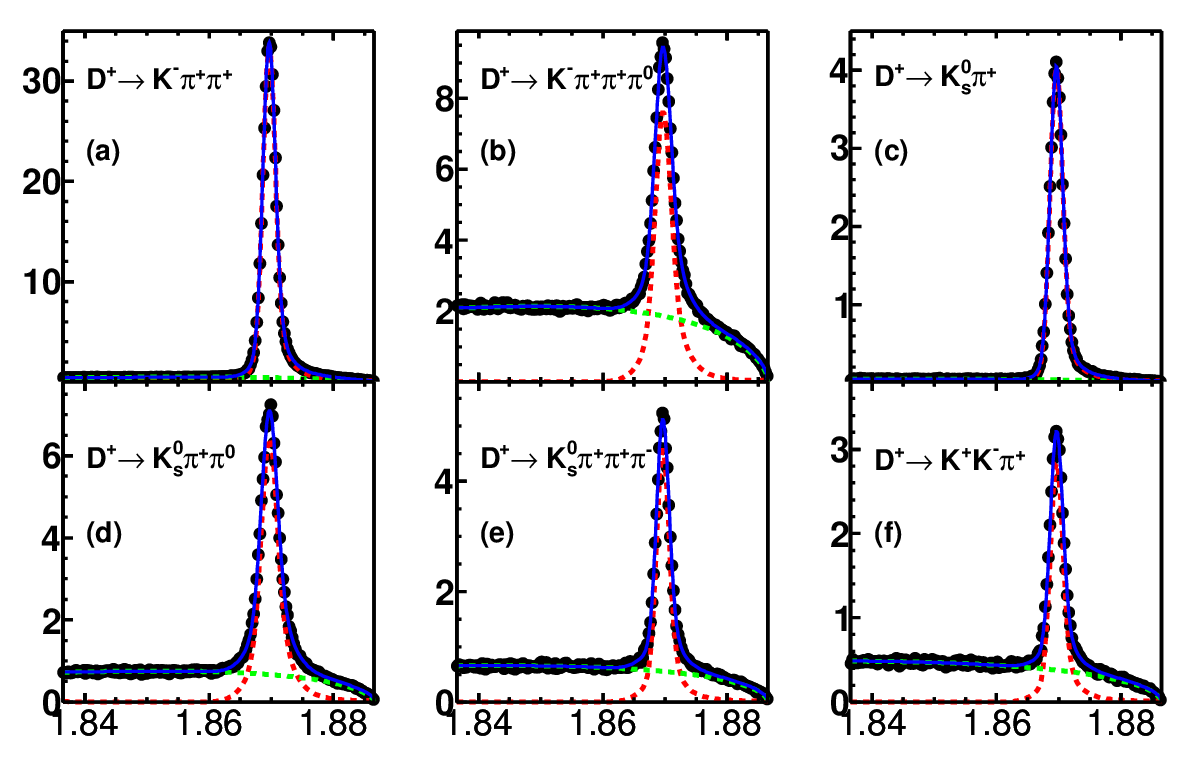}
	\put(39,-2){$M_{\rm BC}$(GeV/$c^2$)}
	\put(-3,8){\rotatebox{90}{Events/(0.25MeV/$c^{2}$) ($\times10^{3}$)}}
	\end{overpic}
  }
  \caption{$M_{\rm BC}$ fit in data for ST decay modes (a) $D^+\to K^-\pi^+\pi^+$, (b) $D^+\to K^-\pi^+\pi^+\piz$, (c) $D^+\to \Ks\pi^+$, (d) $D^+\to \Ks\pi^+\piz$ (e) $D^+\to \Ks\pi^+\pi^+\pi^-$, (f) $D^+\to K^+K^-\pi^+$. The red and green dashed lines are signal and background, respectively. The blue solid lines are the sum of signal and background. The black points are data.
  }
  \label{fig:SingleTagMbcFit}
\end{figure}

To determine the yields of $D^{\pm}$ meson for each ST decay modes, binned maximum-likelihood fits are performed to the $M_{\rm BC}$ distributions in the range from 1.8365 to 1.8865~$\gevcc$, as illustrated in Fig.~\ref{fig:SingleTagMbcFit} for the $\Dp$ meson. In the fit, the signal is modeled by the MC-simulated shape convolved with a double-Gaussian function to take the resolution difference between data and MC simulation into account. The means and widths of the double-Gaussian function are free independent parameters in the fit. The combinatorial background is described by an ARGUS function~\cite{1990argus}. 
Candidate events within $M_{\rm BC}\in (1.863, 1.877)~\gevcc$ are kept for further analysis. The corresponding yields of the ST $D^\pm$ mesons are determined by integrating the fitted signal lineshape in the same $M_{\rm BC}$ range, as summarized in Table~\ref{tab:STyield_STDTeff}. Summing over all six ST modes, the total yields  $N_{\rm ST}^{\rm tot}$ are $(758.2\pm1.4)\times 10^3$ for $\Dp$ mesons and $(763.9\pm1.5)\times 10^3$ for $\Dm$ mesons. 
The detection efficiency of ST reconstruction for the decay mode $i$, $\epsilon^{i}_{\rm ST}$, is obtained from fits to the corresponding $M_{\rm BC}$ distribution of inclusive MC samples, as summarized in Table~\ref{tab:STyield_STDTeff}.

\begin{table}[htbp]
  \begin{center}
  \footnotesize
  \caption{Summary of ST yields $N_{\rm ST}^{i}$, ST efficiencies $\epsilon_{\rm ST}^{i}$(\%), and DT efficiencies $\epsilon_{\rm DT}^{i,c1}$ and $\epsilon_{\rm DT}^{i,c2}$(\%), for the different ST decay modes, which are used to calculate $N_{\rm ST}^{\rm tot}$ and $\epsilon_{\rm sig}$ in Eq.~\ref{eq:BR}. $\epsilon_{\rm DT}^{i,c1}$ and $\epsilon_{\rm DT}^{i,c2}$ are DT efficiencies for signal channels with $\Delta|B-L|=0$ and $\Delta|B-L|=2$, respectively.}
  \begin{tabular}{l|c|c|c|c}
    \hline \hline
    ST modes                          &  $N_{\rm ST}^{i}(\times 10^3)$ &  $\epsilon_{\rm ST}^{i}$ & $\epsilon_{\rm DT}^{i,c1}$ &  $\epsilon_{\rm DT}^{i,c2}$  \\  \hline
    $D^+\to K^-\pi^+\pi^+$            &  $390.2\pm1.1$         &  $50.22\pm0.03$  & 9.79 & 9.26     \\
    $D^+\to K^-\pi^+\pi^+\piz$      &  $124.0\pm0.6$         &  $26.40\pm0.05$   & 5.51 &  5.21  \\
    $D^+\to K^0_{S}\pi^+$               &  ~$45.9\pm0.2$            &  $50.58\pm0.10$    & 9.69 & 9.17  \\
    $D^+\to K^0_{S}\pi^+\piz$         &  $106.7\pm0.6$            &  $27.07\pm0.06$   & 5.66 & 5.36 \\
    $D^+\to K^0_{S}\pi^+\pi^+\pi^-$ &  ~$56.9\pm0.4$            &  $28.16\pm0.08$    & 5.64 & 5.34  \\
    $D^+\to K^+K^-\pi^+$             &  ~$34.6\pm0.3$            &  $41.13\pm0.15$     & 7.88 & 7.46   \\
    \hline
    $D^-\to K^+\pi^-\pi^-$             &  $392.4\pm0.7$         &  $51.19\pm0.03$     & 9.21 & 9.72 \\
    $D^-\to K^+\pi^-\pi^-\piz$      &  $127.7\pm1.0$          &  $26.86\pm0.06$     & 5.19 & 5.47   \\
    $D^-\to K^0_{S}\pi^-$               &  ~$45.5\pm0.2$            &  $50.64\pm0.09$      & 9.12 & 9.62 \\
    $D^-\to K^0_{S}\pi^-\piz$        &  $107.6\pm0.6$            &  $27.21\pm0.05$        & 5.33 & 5.63 \\
    $D^-\to K^0_{S}\pi^-\pi^-\pi^+$ &  ~$56.2\pm0.4$           &  $27.87\pm0.07$       & 5.31 &  5.60 \\
    $D^-\to K^+K^-\pi^-$             &  ~$34.6\pm0.3$           &  $40.40\pm0.12$      & 7.42 & 7.83  \\
    \hline\hline
  \end{tabular}
  \label{tab:STyield_STDTeff}
  \end{center}
\end{table}

\subsection{Double-Tag Events}
\label{sec:DT}
DT signal candidates are selected from the sample of ST $\Dpm$ events by requiring an electron candidate
and no additional charged tracks in the event. 
A two-constraint (2C) kinematical fit is performed by imposing energy and momentum conservation, and constraining the invariant mass of ST $\Dpm$ candidates as well as the mass of the electron--(anti-)neutron system to be the known mass of $D^\mp$ meson~\cite{PDG}, in which the (anti-)neutron is regarded as a missing particle with unknown mass. 
The fit is required to converge, but no further selection on the $\chisq$ of the fit is applied.
The momentum and invariant mass of the (anti-)neutron obtained from the kinematic fit \mbox{($p_{n/\bar{n}}$ and $M_{n/\bar{n}}$)} are recorded for the subsequent analysis. 
To suppress backgrounds, candidate events are required to possess a shower in the EMC around the fitted direction of the \mbox{(anti-)}neutron within an opening angle of $30^{\circ}$. If there are several showers in this region, the one with the largest energy is selected. 

\begin{figure}[!htp]
  \centering
  \mbox{
	\begin{overpic}[width=0.45\textwidth]{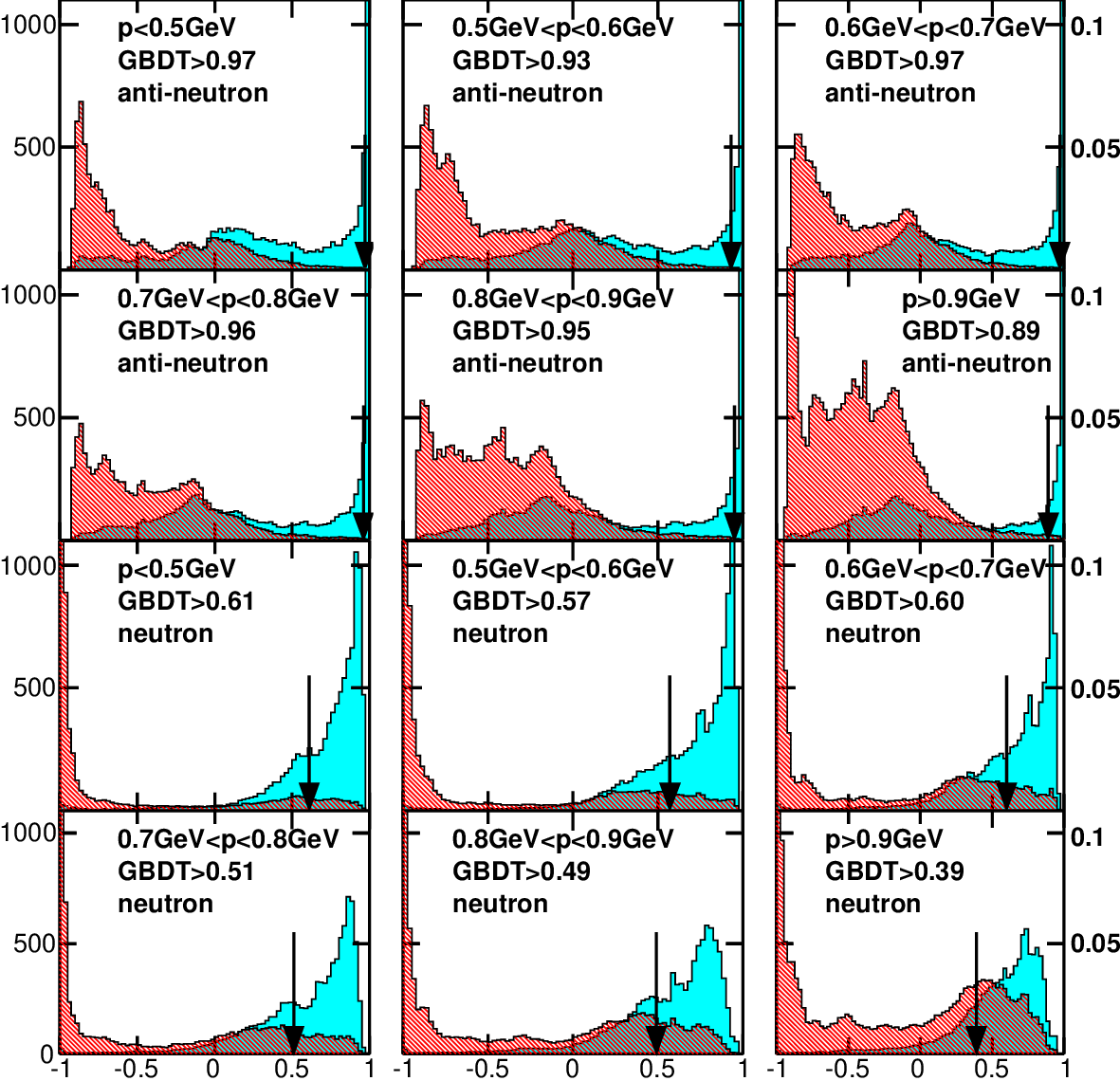}\end{overpic}
	\put(-140,-8){GBDT value}
	\put(0,180){\rotatebox{270}{Signal efficiency (blue)/(0.025)}}
	\put(-240,55){\rotatebox{90}{Background events (red)/(0.025)}}
  }
  \caption{
     The distribution of GBDT values for individual (anti-)neutron momentum bins. The top six histograms are for anti-neutrons and the bottom six are for neutrons. The blue filled histograms are signal and the red hatched ones are backgrounds. The black arrows show the GBDT selection requirement value. The Y axes on the left (right) mean B ($\epsilon$) for background (signal) in the quantity $\epsilon/(1.5+\sqrt{{B}})$.
  }
  \label{fig:BDTGcut}
\end{figure}

MC studies with a generic event-type analysis tool, TopoAna~\cite{TopoAna}, indicate that the backgrounds in the selected samples are dominated by semi-leptonic decays of $D^\pm$ meson with $\Kl$ and $\piz$ mesons in the final state.
Taking into account the result of the MC simulation, the selected showers in the EMC are further required to lie no more than $10^{\circ}$  ($15^{\circ}$)  from the direction of the neutron (anti-neutron) candidate.
In addition, a Multivariate Data Analysis (MVA) based on the shower shape in the EMC is performed based on a Gradient Boosted Decision Trees (GBDT) algorithm. The utilized variables include the total deposited energy $E_{\rm tot}$, the number of hit crystals in the EMC $N_{\rm hit}$, and the A20 and A42 Zernike moments as defined in Ref.~\cite{ZernikeMomentum}. 
In order to train and test the MVA, high-purity training and testing samples, including the (anti-)neutron signal, as well as $\Kl$ and photon backgrounds, are obtained from data using a selection that is independent of any EMC information. The (anti-)neutron sample comes from the decay process $J/\psi\to \bar{p}n\pi^+ + c.c.$, the $\Kl$ sample from $J/\psi\to K\pi\Kl$ and the photon sample from $J/\psi\to\rho\piz$ with $\piz\to\gamma\gamma$.
Studies show that the distributions of the shower-shape variables have a significant dependence on the momentum of the (anti-)neutron.
Therefore, the MVA is performed in separate (anti-)neutron momentum bins of width 100~MeV/$c$.
For a specific (anti-)neutron momentum bin, the training and testing background samples are reweighted according to their expected momentum lineshapes, which are obtained from the inclusive MC samples. 
The distributions of GBDT values in the different momentum bins for the anti-neutron and neutron as well as for the backgrounds are shown in Fig.~\ref{fig:BDTGcut}. 
The selection on the GBDT values, optimized by maximizing the quantity $\epsilon/(1.5+\sqrt{{B}})$~\cite{punzifom}, where $\epsilon$ is the relative efficiency in the MVA signal sample and $B$ is the number of background events normalized to match the luminosity of data in the inclusive MC sample, are applied and shown in Fig.~\ref{fig:BDTGcut}.

The detection efficiencies for finding a matching shower, and for the selection on the opening angle and GBDT value are evaluated from data using the large and high purity control sample of $J/\psi\to pn\pi$ decays. 
The efficiencies of finding a matching shower and for the requirement on the opening angle are studied as a function of two variables: the (anti-)neutron momentum and $\cos\!\theta$, following the procedure described in  Ref.~\cite{ShowerCorrecting}. 
The above efficiencies are directly applied to the signal MC sample with a sampling approach.  

\subsection{Signal Extraction and Fitting}
\label{sec:SignalExtraction}

\begin{figure}[!htp]
  \centering
  \mbox{
	\begin{overpic}[width=0.23\textwidth]{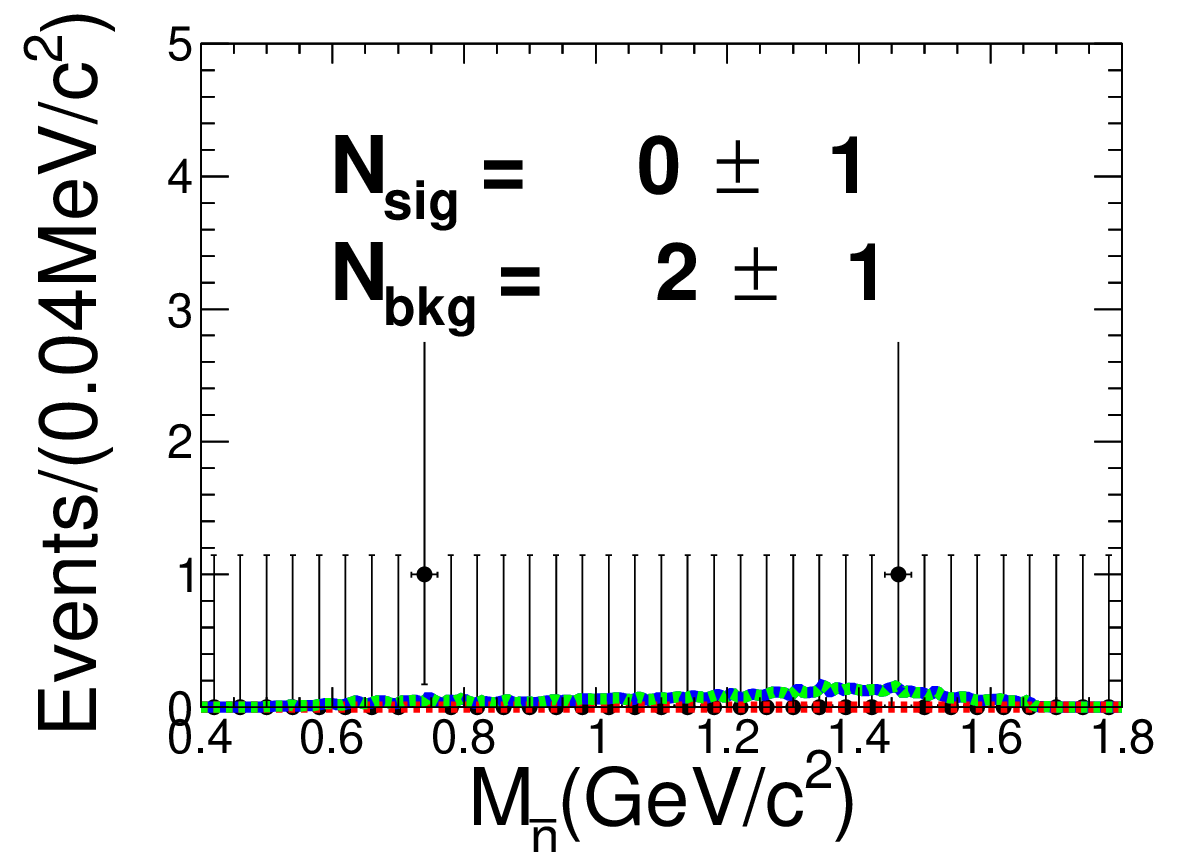}\put(80,60){$(a)$}\end{overpic}
	\begin{overpic}[width=0.23\textwidth]{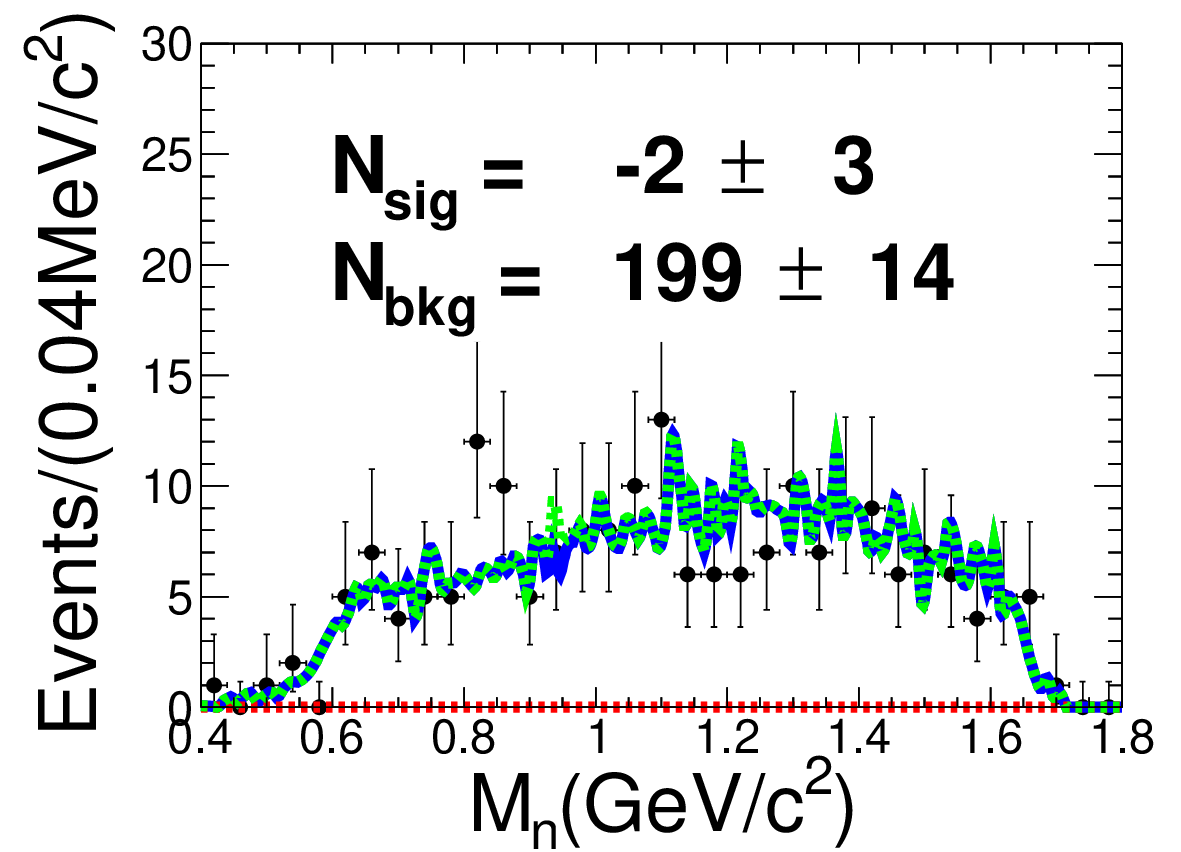}\put(80,60){$(b)$}\end{overpic}
  }
  \mbox{
	\begin{overpic}[width=0.23\textwidth]{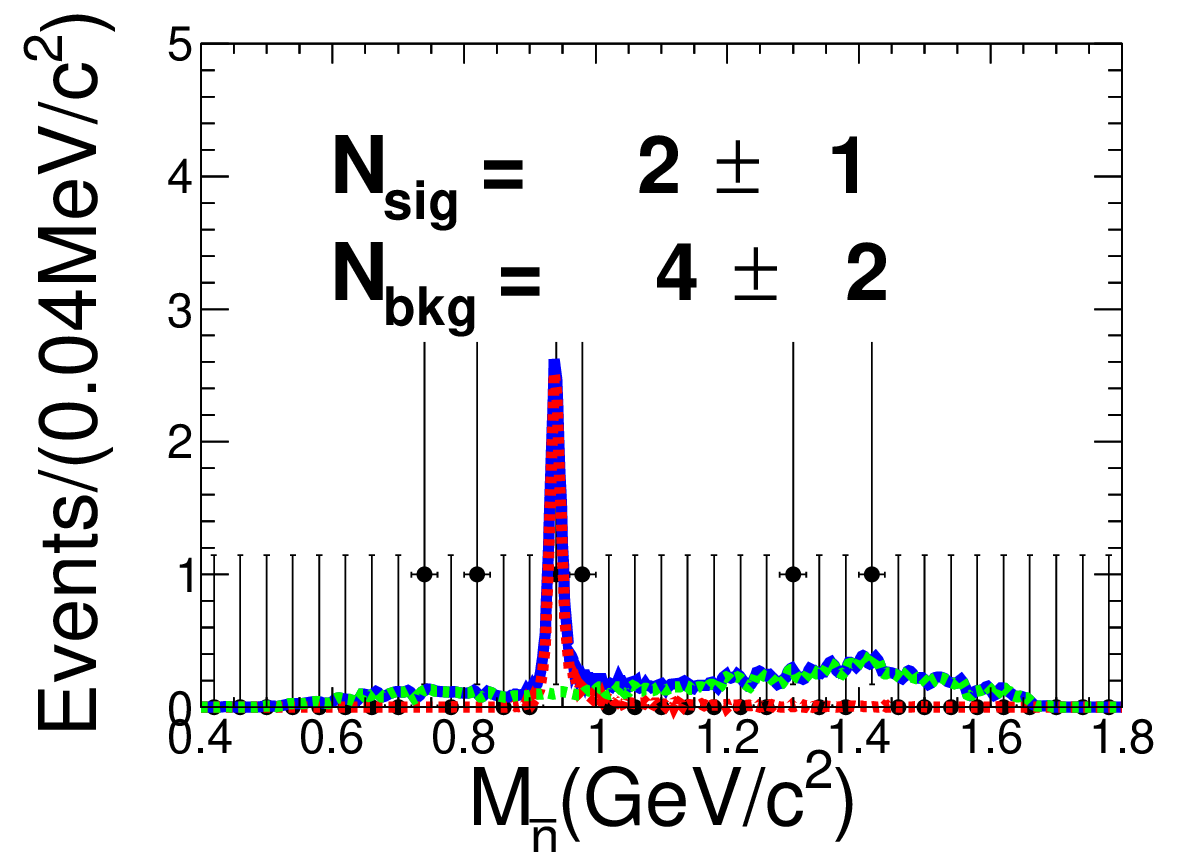}\put(80,60){$(c)$}\end{overpic}
	\begin{overpic}[width=0.23\textwidth]{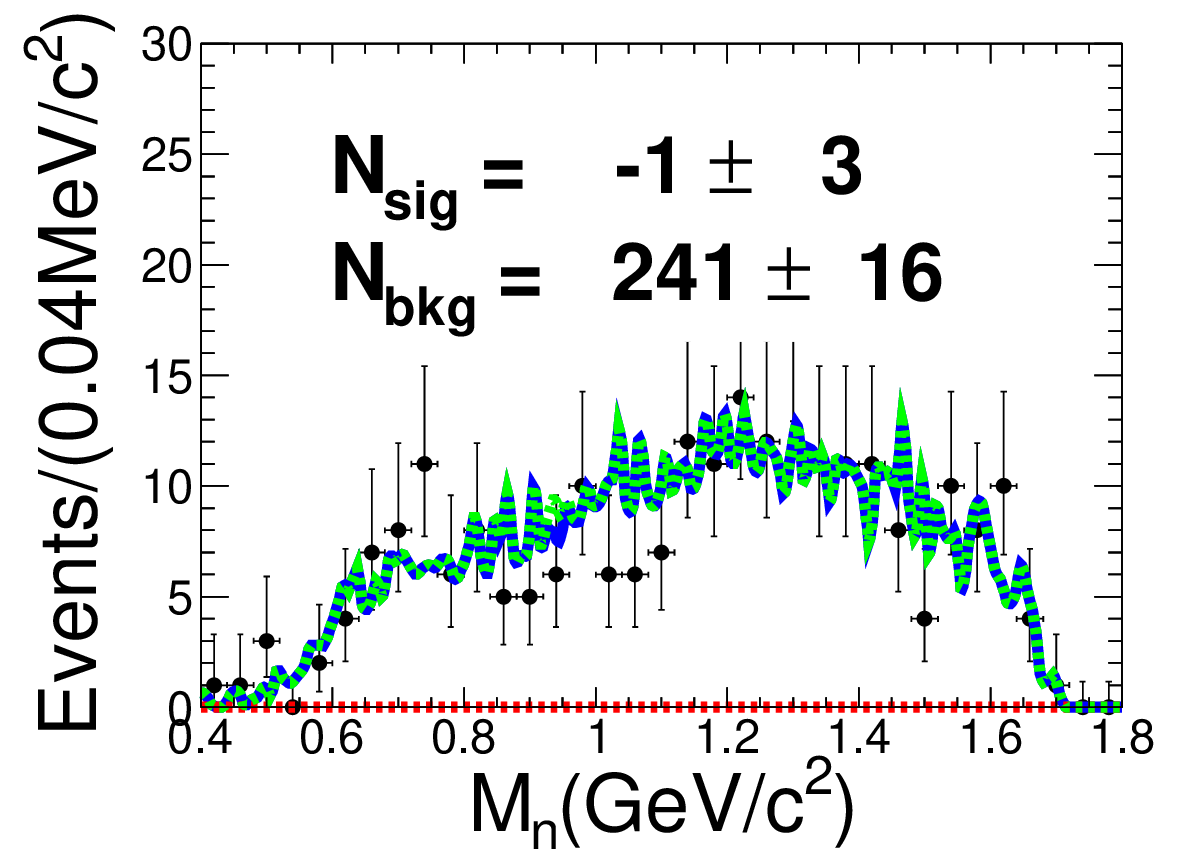}\put(80,60){$(d)$}\end{overpic}
  }
  \caption{
     Fit for $M_{n/\bar{n}}$ distributions for processes (a) $\Dptonbare$, (b) $\Dmtone$, (c) $\Dmtonbare$ and (d) $\Dptone$. The black dots with error bar are data. The red dotted, green dotted and blue solid lines are signal, background,  and the sum of signal and background, respectively.
  }
  \label{fig:Mn_fit}
\end{figure}

The mass distributions of (anti-)neutron $M_{n/\bar{n}}$  from the kinematic fit, after all selection cuts, are shown in Fig.~\ref{fig:Mn_fit} for the four decay processes, where the upper two plots are for the processes $\Dptonbare$ and $\Dmtone$ with $\Delta|B-L|=0$, and the lower two plots are for the processes $\Dmtonbare$ and $\Dptone$ with $\Delta|B-L|=2$. 
No obvious signal is observed.
The DT detection efficiencies for the different ST modes  
are determined to be 
$18.65\%$ for $\Dptonbare$, $19.92\%$ for $\Dmtone$, $19.68\%$ for $\Dptone$ and $18.85\%$ for $\Dmtonbare$.

An unbinned maximum-likelihood fit is performed to each $M_{n/\bar{n}}$ distribution as shown in Fig.~\ref{fig:Mn_fit}, individually. In the fit, signal and background are modeled by the MC-simulated shapes obtained from signal and inclusive MC samples, respectively.  
The yields of signal and background are left free in the fit and the returned values are shown in the plots.
Since no significant signal is observed, conservative ULs are set by combining the two processes with $\Delta|B-L|=0$ and those with $\Delta|B-L|=2$, respectively, as described below.

\subsection{Assignment of Systematic Uncertainties}
The systematic uncertainties related to the efficiency for reconstructing the tag side cancel due to the DT method.
The sources of possible systematic bias that remain include those associated with the DT-side selection efficiency, and the ST and DT yields extraction.

The uncertainties associated with the DT-side event-selection efficiency include those from the electron tracking and PID efficiencies, from the 2C kinematic fit, from the efficiency of finding a matched shower, and from requirements on the angle and the GBDT values. 

The uncertainties on the tracking and PID efficiencies for electrons and positron are studied in control samples, as described in Ref.~\cite{ETrackingPIDUncertainty}. These efficiencies are studied in two-dimensional bins of momentum versus $\cos\!\theta$  for data and MC simulation, individually.
The average relative differences on the efficiencies between data and MC simulation, which is calculated by weighting the corresponding values according to the distribution of the electron/positron signal, are assigned as the systematic uncertainties.

The uncertainty associated with the 2C kinematic fit is studied with a high-purity control sample $D\to \Ks e\nu_{e}\to\pi\pi e\nu_e$ decays  by mimicking the $\Ks \nu_e$ as a missing system. 
The same kinematic fit is performed on this sample, and the efficiency of decays surviving this fit is measured and compared with the corresponding efficiency in MC simulation.
This comparison is made as a function of the invariant mass of the $\Ks \nu_e$ system, and the relative difference around the neutron mass is taken as the systematic uncertainty.

The detection efficiencies for finding a matching shower and for the angle requirement in the EMC for \mbox{(anti-)}neutron are estimated from a control sample of $J/\psi\to pn\pi$ decays in two-dimensional bins of momentum versus $\cos\!\theta$, and then applied directly to the signal MC samples with a sampling approach. 
The dominant source of potential bias from this approach arises from the different physics environment in the EMC between the signal and control sample,  as well as the statistical uncertainty associated with the size of the control sample. 
To estimate the size of this potential bias, we compare the efficiencies between the signal MC sample and the MC-simulated control sample $J/\psi\to pn\pi$, and assign the  small differences observed  as the systematic uncertainties.
The uncertainty associated with the sample size is determined by standard error propagation.

The efficiency of the requirement on the GBDT value is determined from the control sample of $J/\psi\to pn\pi$ decays in different \mbox{(anti-)}neutron momentum bins. Two sources of potential bias are considered: background in the control sample and the choice of momentum binning.
The amount of possible contamination is estimated by fitting the missing mass in the control sample in the different momentum bins, and its full effect is determined on the efficiency measurement and taken as the corresponding systematic efficiency.
The possible bias  associated with the momentum binning is evaluated by varying the bin size and taking the maximum change in the measured efficiency as the systematic uncertainty.

The uncertainties associated with DT selection efficiency are summarized in Table~\ref{tab:Uncertainty}, and the total uncertainties are the quadratic sum of the individual values.

\begin{table}[htbp]
  \centering
  \caption{Summary of the systematic uncertainties on the DT efficiencies (\%).}
  \label{tab:Uncertainty}
  \begin{tabular}{c|c|c|c|c}
    \hline\hline
    Source  & $\Dptonbare$ & $\Dptone$ & $\Dmtonbare$ & $\Dmtone$  \\ \hline
    e tracking  &  0.50  &  0.50  &  0.50  &  0.50  \\
    e PID  &  0.10  &  0.10  &  0.10  &  0.10  \\
    2C fit  &  1.00  &  1.00  &  1.00  &  1.00  \\
    Find shower  &  1.10  &  4.06  &  1.05  &  4.99  \\
    Angle match  &  2.12 &   1.79  &  2.35 &  1.63  \\
    GBDT cut  &  2.16  &  2.33  &  2.16  &  2.33  \\ \hline
    Total         &  3.41 & 5.14  &  3.54   &  5.85  \\
     \hline\hline
  \end{tabular}
\end{table}

The uncertainties on the ST yields are estimated to be $0.5\%$ by studying the variation in results when using a different fit range of (1.8415,1.8865) $\gevcc$, describing the combinatorial background with a  3-order polynomial  rather than an ARGUS function, and by imposing a different endpoint of 1.8863 or 1.8867 $\gevcc$ for the ARGUS function~\cite{STYieldUncertainty}.

The uncertainty associated with the  fit of the $M_{n/\bar{n}}$ distribution arises from the imperfect knowledge of the background shape and the choice of fit range, which will be considered in the determination of the upper limits.

\subsection{Determination of the Upper Limits}

Since no signal is observed, an UL is set on $\Delta|B-L|=0$ processes by performing a fit to the $M_{n/\bar{n}}$ distributions of $\Dptonbare$ and $\Dmtone$, similar to that of Sec.~\ref{sec:SignalExtraction}, but with the signal yields fixed.  A UL is also set on $\Delta|B-L|=2$ processes from a fit to the $\Dmtonbare$ and $\Dptone$ distributions.

\begin{figure}[!htp]
  \centering
	\begin{overpic}[width=0.4\textwidth]{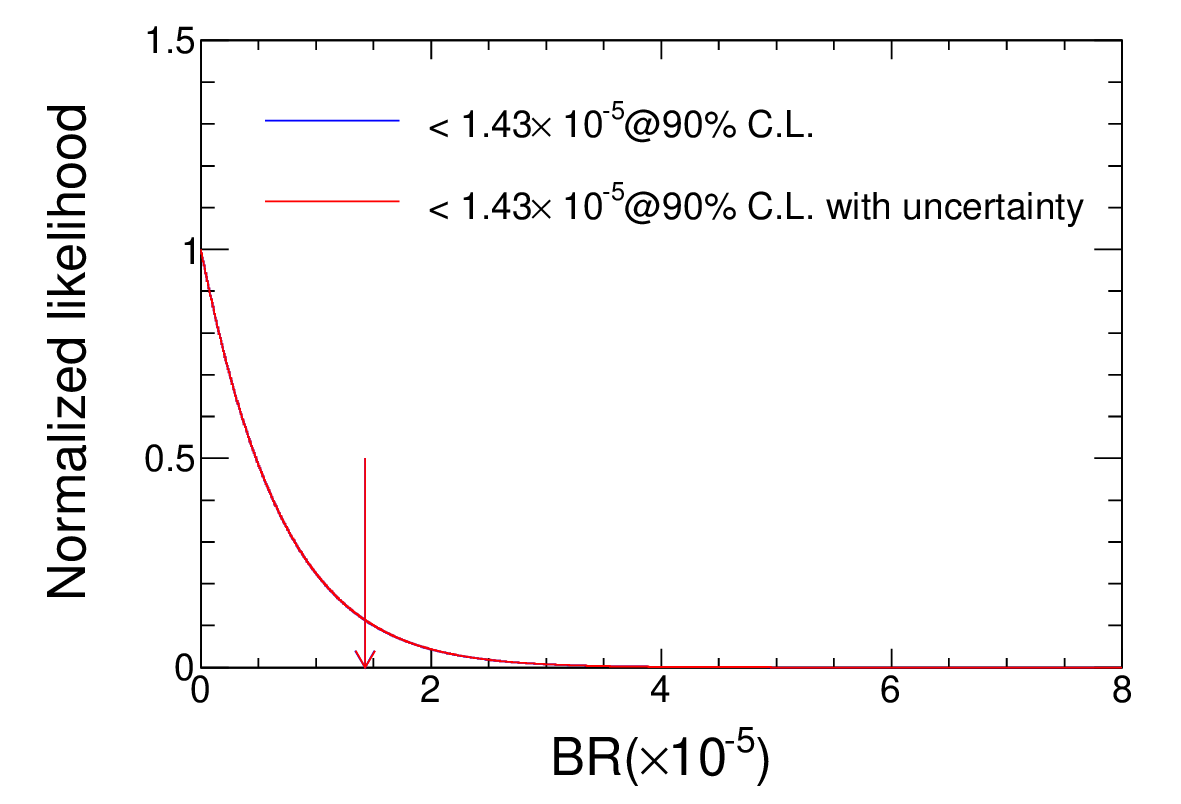}\put(80,55){$(a)$}\end{overpic}
	\begin{overpic}[width=0.4\textwidth]{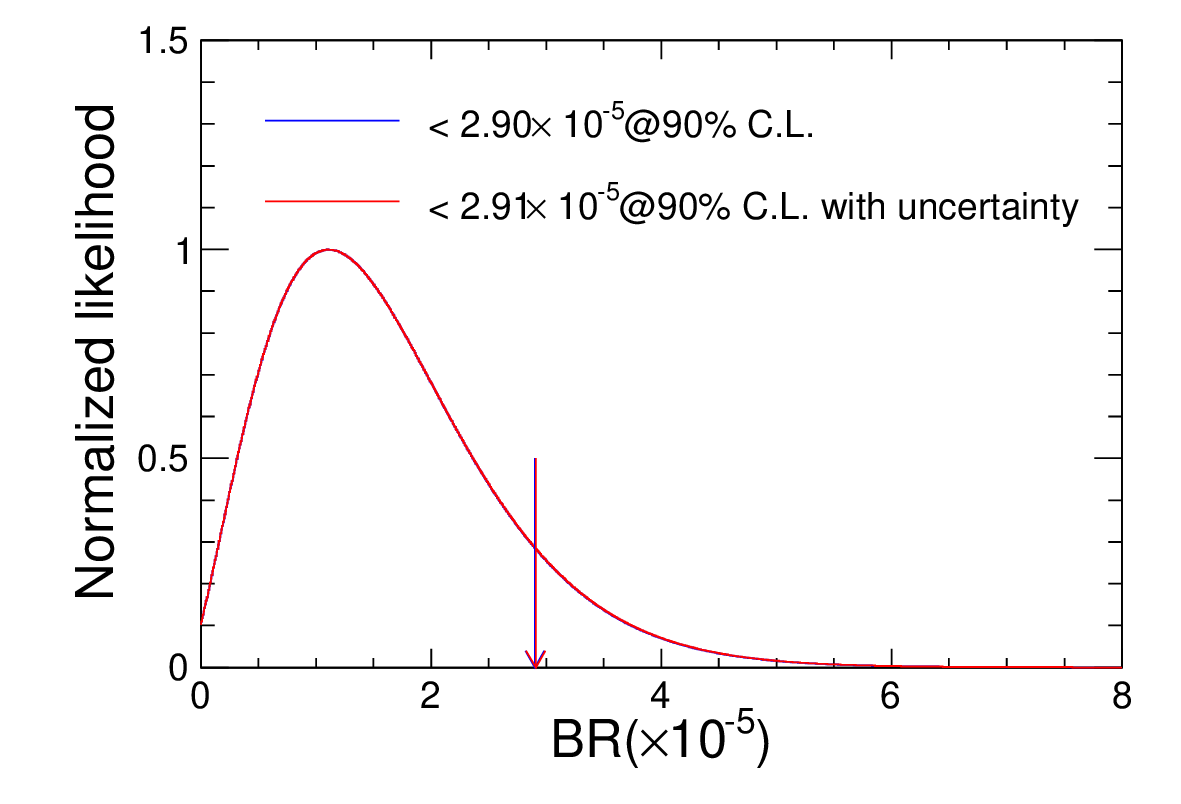}\put(80,55){$(b)$}\end{overpic}
  \begin{overpic}[width=0.4\textwidth]{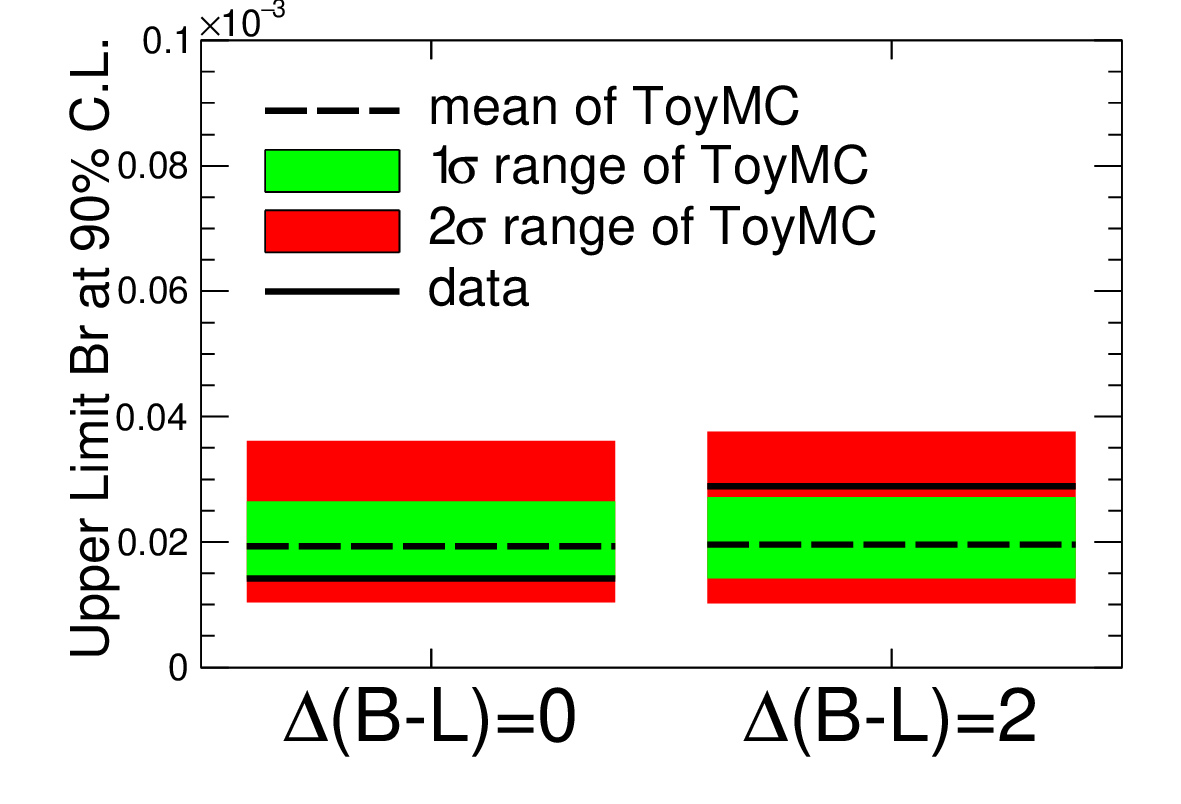}\put(80,55){$(c)$}\end{overpic}
  \caption{
  Likelihood distributions and ULs results in data and Toy MC study. The top and mid figures show the likelihood distribution versus branching fraction for the processes (a) $\Delta|B-L|=0$ and (b) $\Delta|B-L|=2$. The red (blue) lines are the smeared (original) distributions. The red (blue) vertical arrows show ULs of branching fractions with (without) the systematic uncertainty included. Figure (c) shows the branching fractions of data and ToyMC studies. The black solid and dashed lines are ULs from data and means of Toy MC study, respectively. The green (red) ranges show $1(2)$ times of standard deviations interval in Toy MC study. The means and standard deviations are obtained by fitting the UL distribution in ToyMC samples with Gaussian function in log scale. 
  }
  \label{fig:BrResult}
\end{figure}

The fixed signal yields in the fits correspond to  different branching fraction assumptions, according to the effective detection efficiencies, ST yields and the uncertainties. 
Likelihood values are obtained as a function of the branching fraction,
where the effects of systematic uncertainties associated with DT efficiencies and ST yields are included by convolving the likelihood distribution with Gaussian functions of mean zero and width equal to their absolute uncertainties, as described in Refs.~\cite{LikelihoodSmear}.
The ULs on the branching fraction at the 90\% CL are calculated by integrating the likelihood distribution from zero to $90\%$ of the total curve. 
To take into account the effects on the imperfect knowledge of background shape and the fit range on the $M_{n/\bar{n}}$ distributions, alternative fits are performed with different assumptions for  background lineshape and fit range. The most conservative ULs obtained at the $90\%$ CL are shown in Fig.~\ref{fig:BrResult}, which are taken as the final results.
These ULs are $\mathcal{B}_{\Dptonbare}<\BrDataCombA$ and $\mathcal{B}_{\Dptone}<\BrDataCombB$. Studies of ensembles of simulated experiments (`toy MC') that contain no signal give results that are consistent with these measurements within $2\sigma$.

\section{Summary}
In summary, by analyzing $e^{+}e^{-}$ collision data with an integrated luminosity of $\dataIntLumi$  at $\sqrt{s}=3.773$ GeV taken with the BESIII detector, we search for the BNV decays $\Dptonbare$ with $\Delta|B-L|=0$ and $\Dptone$ with $\Delta|B-L|=2$ for the first time. 
No signal is found and the ULs on branching fraction are determined to be $\mathcal{B}_{\Dptonbare}<\BrDataCombA$ and  $\mathcal{B}_{\Dptone}<\BrDataCombB$ for the processes with $\Delta|B-L|=0$ and $\Delta|B-L|=2$ , respectively. 
More data at this collision energy is being collected, up to an integrated luminosity of around 20~fb$^{-1}$~\cite{BESIIIWhithPaper}.  With this larger sample, and assuming no signal, it will be possible to improve the ULs by around a factor of three.

\section*{ACKNOWLEDGMENTS}
The BESIII collaboration thanks the staff of BEPCII and the IHEP computing center for their strong support. This work is supported in part by National Key R\&D Program of China under Contracts Nos. 2020YFA0406400, 2020YFA0406300; National Natural Science Foundation of China (NSFC) under Contracts Nos. 11635010, 11735014, 11835012, 11935015, 11935016, 11935018, 11961141012, 12022510, 12025502, 12035009, 12035013, 12192260, 12192261, 12192262, 12192263, 12192264, 12192265,~11335008, 11625523, 11705192, 11950410506, 12061131003, 12105276, 12122509; the Chinese Academy of Sciences (CAS) Large-Scale Scientific Facility Program; Joint Large-Scale Scientific Facility Funds of the NSFC and CAS under Contract Nos. U1832207, U1732263, U1832103, U2032111; CAS Key Research Program of Frontier Sciences under Contract No. QYZDJ-SSW-SLH040; 100 Talents Program of CAS; The Institute of Nuclear and Particle Physics (INPAC) and Shanghai Key Laboratory for Particle Physics and Cosmology; ERC under Contract No. 758462; European Union's Horizon 2020 research and innovation programme under Marie Sklodowska-Curie grant agreement under Contract No. 894790; German Research Foundation DFG under Contracts Nos. 443159800, Collaborative Research Center CRC 1044, GRK 2149; Istituto Nazionale di Fisica Nucleare, Italy; Ministry of Development of Turkey under Contract No. DPT2006K-120470; National Science and Technology fund; National Science Research and Innovation Fund (NSRF) via the Program Management Unit for Human Resources \& Institutional Development, Research and Innovation under Contract No. B16F640076; STFC (United Kingdom); Suranaree University of Technology (SUT), Thailand Science Research and Innovation (TSRI), and National Science Research and Innovation Fund (NSRF) under Contract No. 160355; The Royal Society, UK under Contracts Nos. DH140054, DH160214; The Swedish Research Council; U. S. Department of Energy under Contract No. DE-FG02-05ER41374.


\begin{thebibliography}{99}
\bibitem{sakharov}
A. D. Sakharov,
  \href{http://jetpletters.ru/ps/1643/article_25089.shtml}{JETP Lett. \textbf{5}, 24 (1967).}
\bibitem{GUTs1}
  J. C. Pati and A. Salam,
  \href{https://doi.org/10.1103/PhysRevD.8.1240}{\PRD {\bf 8}, 1240 (1973).}
\bibitem{GUTs2}
  H. Georgi and S. L. Glashow,
  \href{https://doi.org/10.1103/PhysRevLett.32.438}{\PRL {\bf 32}, 438 (1974).}
\bibitem{GUTs3}
  S. Raby,
  \href{https://arxiv.org/abs/hep-ph/0608183}{arXiv:hep-ph/0608183v1.}
\bibitem{GUTs4}
  A. Davidson,
  \href{https://journals.aps.org/prd/abstract/10.1103/PhysRevD.20.776}{\PRD {\bf 20}, 776 (1979).}
\bibitem{GUTs5}
  R. E. Marshak and R. N. Mohapatra,
  \href{https://doi.org/10.1016/0370-2693(80)90436-0}{\PLB {\bf 91}, 222 (1980).}
\bibitem{SUSY1}
   S. Lola and G. G. Ross,
   \href{https://doi.org/10.1016/0370-2693(93)91246-J}{\PLB {\bf 314}, 336 (1993).}
\bibitem{SUSY2}
  N. Polonsky,
  \href{https://arxiv.org/abs/hep-ph/0108236}{arXiv:hep-ph/0108236v1}
\bibitem{B-LInProtonDecay}
  F. Wilczek, A. Zee
  \href{https://doi.org/10.1016/0370-2693(79)90475-1}{\PLB {\bf 88}, 311 (1979).}
\bibitem{DtopebyCLEO}
  P. Rubin {\it et al.}  [CLEO Collaboration],
  \href{https://doi.org/10.1103/PhysRevD.79.097101}{\PRD {\bf 79}, 097101 (2009).}
\bibitem{BNVInBbyBARBAR}
  P. del Amo Sanchez {\it et al.} [BABAR Collaboration],
  \href{https://doi.org/10.1103/PhysRevD.83.091101}{\PRD {\bf 83}, 091101(2011).}
\bibitem{BNVInHyperonbyCLAS}
  M. E. McCracken {\it et al.},
  \href{https://doi.org/10.1103/PhysRevD.92.072002}{PRD {\bf 92}, 072002(2015).}
\bibitem{DtoLambdae}
  M. Ablikim {\it et al.}  [BESIII Collaboration],
  \href{https://doi.org/10.1103/PhysRevD.101.031102}{\PRD {\bf 101}, 031102 (2020).}
\bibitem{SUSYmodel}
   W. S. Hou, M. Nagashima, and A. Soddu,
 \href{https://journals.aps.org/prd/abstract/10.1103/PhysRevD.72.095001}{\PRD {\bf 72},  095001(2005).}  
\bibitem{BESIIIDetector}
  M.\ Ablikim {\it et al.} [BESIII Collaboration],
  \href{https://www.sciencedirect.com/science/article/pii/S0168900209023870}{\NIMA {\bf 614}, 345 (2010).}
\bibitem{bepcii}
   C.~H.~Yu {\it et al.},
  \href{https://accelconf.web.cern.ch/ipac2016/doi/JACoW-IPAC2016-TUYA01.html}{Proceedings of IPAC2016, Korea, 2016.}
\bibitem{geant4}
  S.\ Agostinelli {\it et al.} [{\sc geant4} Collaboration],
  \href{https://www.sciencedirect.com/science/article/pii/S0168900203013688}{\NIMA {\bf 506}, 250 (2003).}
\bibitem{BESIIIMC}
  Z.\ Y.\ Deng {\it et al.},
  \href{https://www.scopus.com/record/display.uri?eid=2-s2.0-33744746981&origin=inward&txGid=ff1cc68fdd56b4408abd0ee10ad71ba8}{HEP \& NP {\bf 30}, 371 (2006).}
\bibitem{BOSS}
  W.\ D.\ Li {\it et al.}, 
  \href{https://indico.cern.ch/event/408139/contributions/979815/}{Proceeding of CHEP2006 (Mumbai, India, 13-17 February 2006).}
\bibitem{KKMC}
  S.\ Jadach, B.\ F.\ L.\ Ward, and Z.\ Was,
  \href{https://journals.aps.org/prd/abstract/10.1103/PhysRevD.63.113009}{\PRD {\bf 63}, 113009\ (2001);}
  \href{http://dx.doi.org/10.17632/8w4vrpttpc.1}{Comput.\ Phys.\ Commum.\ {\bf 130},\ 260 (2000).}
\bibitem{EVTGEN}
  D.\ J.\ Lange,
  \href{https://doi.org/10.1016/S0168-9002(01)00089-4}{\NIMA {\bf 462}, 152 (2001).}
  R.\ G.\ Ping, 
  \href{https://iopscience.iop.org/article/10.1088/1674-1137/32/8/001}{\CPC {\bf 32}, 599 (2008).}
\bibitem{PDG}
  P.\ A.\ Zyla {\it et al.} [Particle Data Group],
  \href{https://pdglivetest.lbl.gov/Viewer.action}{Prog.\ Theor.\ Exp.\ Phys.\ {\bf 2020},\ 083C01 (2020).}
\bibitem{LUNDCHARM1}
  J.\ C.\ Chen, G.\ S.\ Huang, X.\ R.\ Qi, D.\ H.\ Zhang, and Y.\ S.\ Zhu,
  \href{https://journals.aps.org/prd/abstract/10.1103/PhysRevD.62.034003}{\PRD {\bf 62}, 034003 (2000).}
\bibitem{LUNDCHARM2}
  R.\ L.\ Yang, R.\ G.\ Ping and H. Chen,
  \href{http://cpl.iphy.ac.cn/10.1088/0256-307X/31/6/061301}{Chin.\ Phys.\ Lett.\ {\bf 31}, 061301 (2014).}
\bibitem{PHOTOS}
  E.\ Richter-W$\text{\c{a}}$s,
  \href{https://doi.org/10.1016/0370-2693(93)90062-M}{\PLB {\bf 303}, 163 (1993).}
\bibitem{DoubleTag}
  R.\ M.\ Baltrusaitis {\it et al.} [MARKIII Collaboration], 
  \href{https://journals.aps.org/prl/abstract/10.1103/PhysRevLett.56.2140}{\PRL {\bf 56}, 2140 (1986)}  
\bibitem{1990argus}
  H.\ Albrecht {\it et al.} [ARGUS Collaboration],
  \href{https://doi.org/10.1016/0370-2693(90)91293-K}{\PLB {\bf 241}, 278 (1990).}
\bibitem{TopoAna}
  X.\ Y.\ Zhou, S.\ X.\ Du, G.\ Li and C.\ P.\ Shen,
  \href{https://doi.org/10.1016/j.cpc.2020.107540}{Comput. Phys. Commun. {\bf 258}, 107540 (2021).}
\bibitem{ZernikeMomentum}
  R. Sinkus and T. Voss,
  \href{https://doi.org/10.1016/S0168-9002(97)00524-X}{\NIMA {\bf 391}, 2 (1997).}
\bibitem{punzifom}
  G. Punzi,
  \href{https://arxiv.org/abs/physics/0308063}{arXiv:physics/0308063.}
\bibitem{ShowerCorrecting}
  L. Liu, X. R. Zhou and H. P. Peng,
  \href{https://arxiv.org/abs/2111.10789}{arXiv:2111.10789 [hep-ex].}
\bibitem{ETrackingPIDUncertainty}
  M.\ Ablikim {\it et al.} [BESIII Collaboration],
  \href{https://doi.org/10.1103/PhysRevLett.127.131801}{\PRL {\bf 127}, 131801 (2021).}
\bibitem{STYieldUncertainty}
  M.\ Ablikim {\it et al.} [BESIII Collaboration],
  \href{https://iopscience.iop.org/article/10.1088/1674-1137/40/11/113001}{\CPC {\bf 40}, 113001 (2016).}
\bibitem{LikelihoodSmear}
  K. Stenson, 
  \href{https://arxiv.org/abs/physics/0605236}{arXiv:physics/0605236.}
\bibitem{BESIIIWhithPaper}
  M.\ Ablikim {\it et al.} [BESIII Collaboration],
  \href{https://iopscience.iop.org/article/10.1088/1674-1137/44/4/040001}{\CPC {\bf 44}, 040001 (2020).}
\end{thebibliography}
\end{document}